\documentclass[a4paper]{article}

\usepackage{INTERSPEECH2019}
\usepackage{multirow}
\usepackage{float,tabularx}
\usepackage{amssymb,amsmath,bm}

\newcommand{\argmax}{\arg\!\max}

\title{Zero resource speech synthesis using transcripts \\ derived from perceptual acoustic units}
\name{Karthik Pandia D S, Hema A Murthy}
\address{
  Indian Institute of Technology Madras}
\email{\{pandia,hema\}@cse.iitm.ac.in}

\begin{document}

\maketitle
\begin{abstract}

Zerospeech synthesis is the task of building vocabulary independent speech synthesis systems, where  transcriptions are not available for training data. It is therefore necessary to convert training data into a sequence of fundamental acoustic units that can be used for synthesis during test.  This paper attempts to discover, and model perceptual acoustic units consisting of steady state, and transient regions in speech.  The transients roughly correspond to  CV, VC units, while the steady-state corresponds to sonorants and fricatives. The speech signal is first preprocessed by segmenting the same into CVC-like units using a short-term energy-like contour. These CVC segments are clustered using a connected components-based graph clustering technique. The clustered CVC segments are initialized such that the onset (CV) and decays (VC) correspond to transients, and the rhyme corresponds to steady-states. Following this initialization, the units are allowed to re-organise on the continuous speech into a final set of AUs in an HMM-GMM framework. AU sequences thus obtained are used to train synthesis models. The performance of the proposed approach is evaluated on the Zerospeech 2019 challenge database. Subjective and objective scores show that reasonably good quality synthesis with low bit rate encoding can be achieved using the proposed AUs.

\end{abstract}

\section{Introduction}

Zero resource speech processing is a sub-field in speech processing which does not use any transcribed data for applications that generally mandate the availability of transcriptions to train models. Such applications include speech recognition, keyword spotting, document classification, text-to-speech synthesis, to name a few. Zero resource speech processing is useful for digital processing of languages that either have low audio resources, or languages that do not have a script.  Additionally, AUs discovered can also give new insights into the nature of  AUs produced/perceived.

Zero resource speech processing problem is generally addressed by acoustic unit discovery (AUD). Since the phonetic transcriptions are not available for the audio, AUD aims to discover a set of acoustic units directly from the audio. The obtained transcriptions, either as tokens or as posterior representation, are used to train models. As only audio is available for training, the problem is an unsupervised clustering problem, also termed as segmentation problem. Clustering similar feature vectors to form speech units is the most common approach \cite{qiao2008unsupervised, zhang2009unsupervised, wang2011unsupervised} . The final acoustic unit models are obtained by modelling the clusters. One of the principal issues in AUD is to identify the initial time scale for processing speech. In most existing approaches, a frame of about 25ms length is chosen as the smallest unit for clustering. The frames are clustered based on a similarity measure such as Euclidean distance; For segmentation problem, nearby frames with high similarity measure are merged based on a bottom-up hierarchical clustering approach \cite{qiao2008unsupervised, wang2011unsupervised}. Bayesian approaches can be either parametric or non-parametric. Parametric approaches \cite{zhang2009unsupervised,wang2011unsupervised} for AUD use expectation-maximization approach to cluster or discover  and model AUs. Non-parametric approaches use variational inference \cite{ondel2016variational} or Gibbs sampling \cite{lee2012nonparametric,torbati2016nonparametric,heck2016unsupervised}.
The approaches can be probabilistic generative modelling followed by ANN based discriminative clustering \cite{zhang2012resource,ebbers2017hidden}. There have been a few top-down approaches proposed in the literature\cite{varadarajan2008unsupervised,jansen2013weak}. In \cite{varadarajan2008unsupervised}, the states in an HMM are iteratively split into multiple states based on a splitting criterion. In \cite{jansen2013weak}, the mixtures of UBM are clustered into a set of AUs.

Even though the approaches start with the frame as the smallest unit, the minimum resolution for the units is usually fixed by a duration constraint. Even though the speech units discovered by any AU techniques will be of the size of syllables or context-dependent phones, most approaches do not capitalise on this idea. Rather, the AU size is determined by the measure of similarity between the frames. There are a few approaches that fix the time scale of the acoustic units according to perceptual units in speech. For instance, Nagarajan's work \cite{nagarajan2004language} on language identification segments speech into syllable-like units using the signal energy. Similarly, R{\"a}s{\"a}nen \cite{rasanen2015unsupervised} used a theta-band oscillator to segment the speech into syllable-like segments. Both the approaches cluster the segments to obtain final acoustic units. Both the approaches use syllabic structure to discover AUs, where the transients are part of the syllable.  The vowel region in a syllable corresponds to 90\% of syllable length,  which in turn leads to poor modeling of transients.

We propose an approach to discover and model acoustic units that are perceptual units. As the task is to synthesise speech using the acoustic units, perceptual units are superior to other kinds of units. Instead of just clustering the syllabic units as proposed by R{\"a}s{\"a}nen and Nagarajan, we proposed AUs that models both the syllabic regions and inter-syllabic regions. The AUs that we model correspond to steady-state and transient regions in speech. To model the AUs, we first train the models on shorter syllable-like segments, which is motivated by children's language acquisition. For this, the syllable-like units are used to initialize the models corresponding to onset, rhyme and coda. After the models are fine-tuned by training on syllable-like segments, the models are retrained on continuous speech to obtain final models. 

\begin{figure*}[t]
  \centering
  \includegraphics[width=1.0\linewidth]{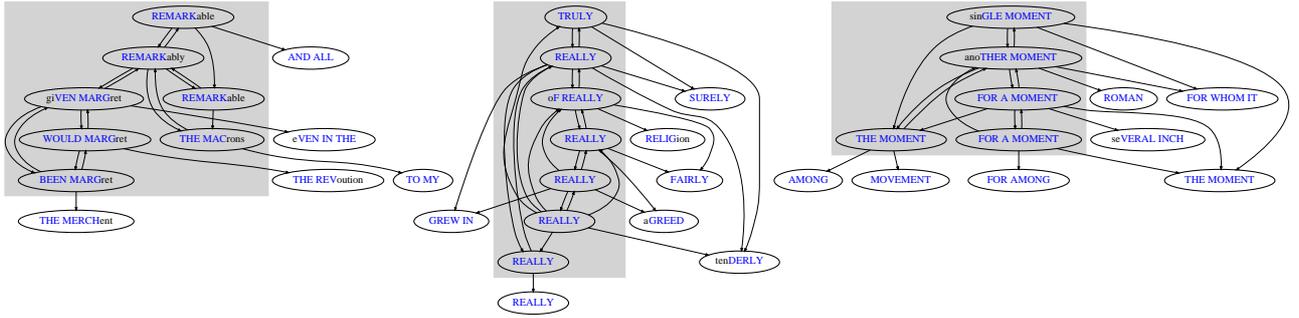}
  \caption{Three different clusters represented in terms of graphs along with the constituent units obtained after the initial clustering}
  \label{fig:cluster}
\end{figure*}

The proposed approach is evaluted on zerospeech 2019 challenge \cite{zerospeech2019} dataset. The results showed in this paper correspond to that of the systems submitted to the challenge. Zerospeech challenge conducted every two years focuses on zero-resource speech processing. The broad objective of the challenge is to construct an end-to-end spoken dialogue system for an unknown language. The first two challenges \cite{versteegh2015zero, dunbar2017zero} focused on unit discovery and lexicon discovery. The third and the current challenge focuses on unit discovery and speech synthesis using the discovered units. Subjective evaluation shows that zero resource speech synthesis is indeed possible and the results are comparable to a supervised system which uses transcribed audio for both acoustic modelling and synthesis. 

The rest of the paper is organised as follows. Section~\ref{sec:proposed} explains the proposed approach for acoustic unit discovery. Section~\ref{sec:dataset} gives a brief overview of the experimental setup and data set used for experimentation. The results are shown and discussed in Section~\ref{sec:results}. Section~\ref{sec:conclusion} concludes the paper.

\section{Proposed AUD technique}
\label{sec:proposed}

As the task is to synthesise speech using the discovered acoustic units, AUs of the form of perceptual units are better than AUs of arbitrary form. Units of the size of syllables are accepted to be the basic units of speech perception. Unlike phonemes and context-dependent phones, approximate syllabic units can be obtained by processing the signal. The peaks and valleys in the envelope of the speech signal naturally segment the speech signal into syllable-like units. Instead of directly clustering such syllabic segments to obtain syllable-like AU, as in case of Nagarajan and R{\"a}s{\"a}nen, the inter-syllabic regions are explicitely modelled. 

\begin{figure}[h]
  \centering
  \includegraphics[width=1.0\linewidth]{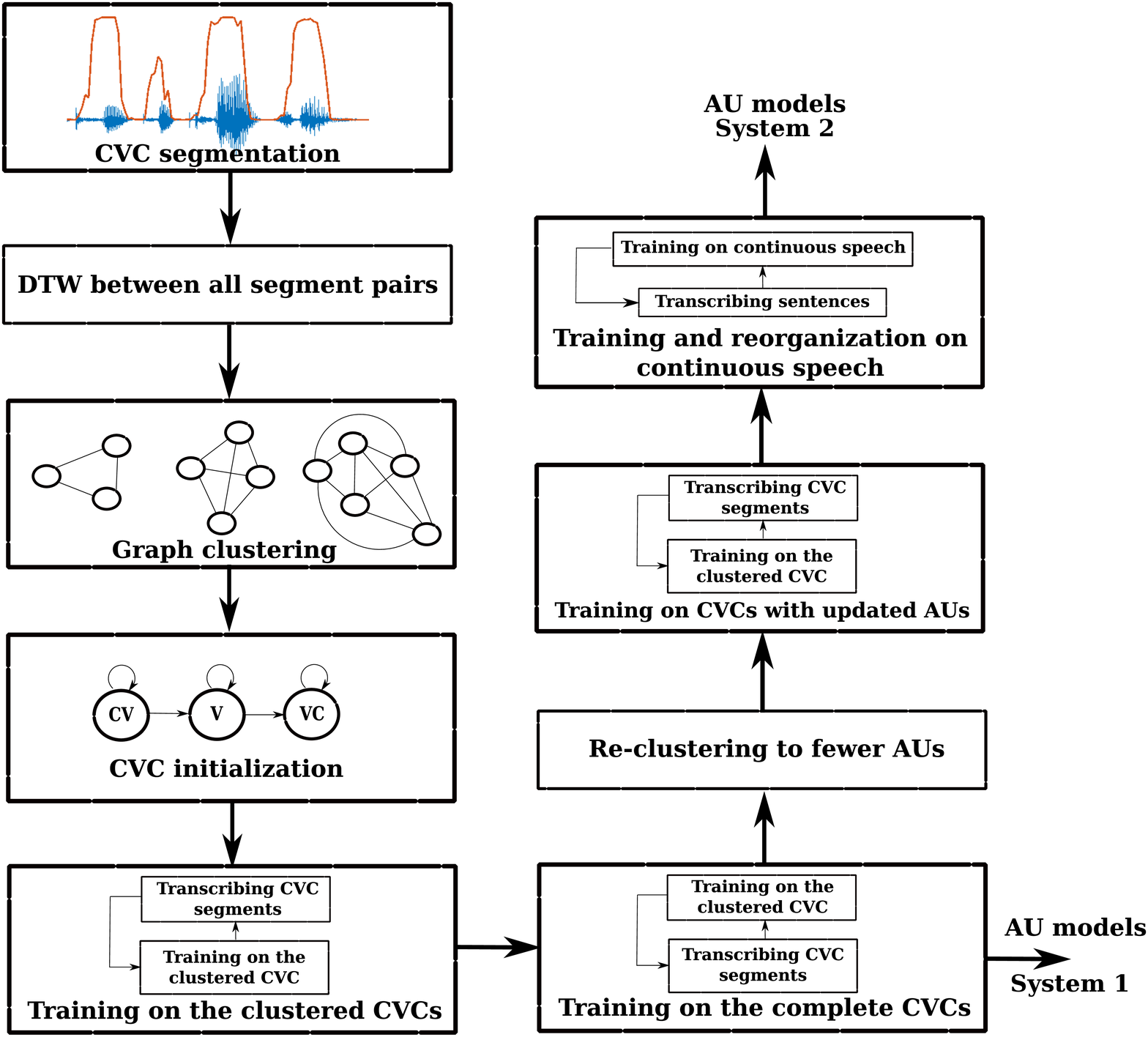}
  \caption{Block diagram of the proposed AUD}
  \label{fig:block}
\end{figure}

The overall flow of the proposed unit discovery technique is shown in Figure~\ref{fig:block}. The speech is first segmented into syllable-like units using a processed short-time energy function. The syllable-like segments are clustered by applying dynamic time warping (DTW) between all possible segment pairs. K-nearest neighbour connected component clustering approach \cite{brito1997connectivity} is applied to cluster similar segments. For each segment, the k-nearest segments are identified by using the DTW scores. A graph is then constructed with segments as nodes. An edge exists between two nodes i and j if i is in the set of the k-nearest neighbours of j and vice-versa. All the segments in a cluster are homogeneous. A snippet of three clusters along with the units is shown in Figure~\ref{fig:cluster}. An arrow between a node A to another node B is present if B is in the k-nearest neighbour of A. The segments in a cluster is sonorant regions and hence can spawn more than two words. The units belonging to each cluster are contained in the corresponding shaded regions. The figure shows that the units within a cluster are highly homogeneous. 

\begin{figure}[h]
 \centering
 \includegraphics[width=1.0\linewidth]{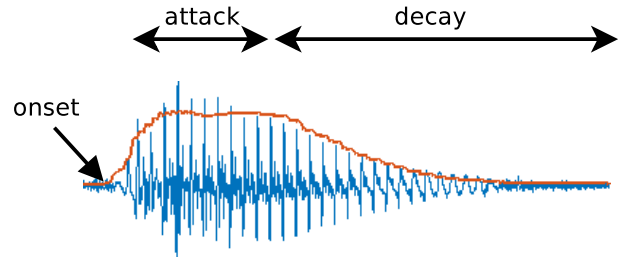}
 \caption{Illustration of onset, attack and decay for a syllable-like segment}
 \label{fig:oad}
\end{figure}

\begin{table*}[t]
\caption{Dataset used for experiments}
\label{tab:data}
\resizebox{\linewidth}{!}{
\begin{tabular}{c|c c c||c c c} \hline
{\bf \multirow{2}{*}{Dataset}} & \multicolumn{3}{c||}{\bf Development language (English)} & \multicolumn{3}{c}{\bf Surprise/Test language (Indonesian)} \\
& \# Speakers & \# Utterances & Duration & \# Speakers & \# Utterances & Duration \\ \hline
Train unit & 100 & 5941 & 15h 40m & 112 & 15340 & 15h \\  \hline
\multirow{2}{*}{Train voice} & 1 male & 970 & 2h & \multirow{2}{*}{1 female} & \multirow{2}{*}{1862} & \multirow{2}{*}{1h 30m} \\
& 1 female & 2563 & 2h 40m & & & \\ \hline
Test & 24 & 455 & 28m & 15 & 405 & 29m \\ \hline
\end{tabular}}
\end{table*}

\begin{table*}[t]
\caption{Results of the proposed , baseline and topline approaches on development and test languages}
\label{tab:result}
\resizebox{\linewidth}{!}{
\begin{tabular}{c|c c c c c||c c c c c} \hline
\bf \multirow{2}{*}{System} & \multicolumn{5}{c||}{\bf Development language (English) } & \multicolumn{5}{c}{\bf Test language (Indonesian)} \\
& MOS & CER & Similarity & ABX & Bitrate & MOS & CER & Similarity & ABX & Bitrate  \\ \hline
Baseline & 2.5	& 0.75	& 2.97	& 35.63	& 71.98 & 2.07	& 0.62 & 3.41	& 27.46	& 74.55 \\
System 1 & 2.82 & 0.55 & 2.76 & 29.66 & 138.59 & 2.53 & 0.43 & 3.58 & 23.56 & 115.43 \\
System 2 & 2.77 & 0.61 & 3 & 28.16 & 92.75 & 2.02 & 0.48 & 3.21 & 20.77 & 94.15  \\
Top-line & 2.77 & 0.44 & 2.99 & 29.85 & 37.73 & 3.92 & 0.28 & 3.95 & 16.09 & 35.2  \\ \hline
\end{tabular}}
\end{table*}

The segments are assumed to contain three distinct units characterized by a vowel onset, rhyme and a vowel offset as illustrated in Figure~\ref{fig:oad}. The steady-state regions are not limited just to vowels, but also sonorant sounds. Hence, the rhyme regions for the segments shown in Figure~\ref{fig:cluster} contain approximant and nasal sounds. The modelling of acoustic units is a two-stage process. In the first stage, the models are trained only on the CVC segments. All the segments in a cluster are assumed to correspond to a sequence of three symbols consisting of a sequence of onset, rhyme and offset \(OS_i, RH_i, OF_i\). The assumed units are used as transcripts for all the segments in a cluster, and Hidden Markov model (HMM) is used to train the individual units. The models are realigned and retrained in a self-training fashion until the overall likelihood of the data with respect to the model converges. Let $O$ correspond to feature vectors and $W_{true}$ corresponds to the true label sequence, and $\Theta$ represents HMM parameters.  Both $\Theta$ and $W_{true}$ are unknown quantities. The initial label sequence is obtained using unsupervised initial clusters. Given the label sequence, the new model parameter, $\Theta_{new}$ is estimated as,
\begin{equation}
	\Theta_{new} = \argmax \limits_\Theta P \left(O, W_{old}|\Theta \right)
\end{equation}
A new label sequence is obtained, given the updates $\Theta_{new}$
\begin{equation}
	W_{new} = \argmax\limits_W P(O, W|\Theta_{new})
\end{equation}

This self-training approach is similar to that of \cite{wang2011unsupervised}. As the duration of the segments is short, this approach is useful to obtain a good initial model although inter-segment transients are not initialized. The obtained AU models are then used to transcribe the continuous utterances of long duration. This will help to train transients in the inter-segment regions. Similar to the first stage, a self-training approach is applied to train models. The obtained model is used in system 1. The number of units obtained by this approach is much larger than the actual perceptual units in any language. Hence, it should be possible to obtain a smaller set of AU. The AUs are merged using a modified k-mean algorithm to obtain a smaller set of 40 units.  Using the new set of labels, the two stages of training is repeated to obtain the models for system 2.

\section{Dataset and experiments}
\label{sec:dataset}
The proposed approach for speech unit discovery followed by speech synthesis is evaluated on the database provided as part of the zerospeech 2019 challenge. Parameter tuning is performed on development data, and the same procedure is repeated on the test data. The language for the development data is English, and the Indonesian language is used as test data. More details about the test corpus can be found at \cite{sakti2008development1, sakti2008development2}. Table~\ref{tab:data} summarizes the dataset used for the experiments. Train unit dataset was provided to perform AUD and acoustic modelling. The duration of train data for both development and test languages are approximately 15 hours with 100 speakers in development data and 112 speakers in test data. The synthesis models are trained on a target voice. For the development data, one male and one female voices were provided, whereas for the test data, one female voice was provided. The audio provided in the test data has to be synthesized in the target speaker's voice. The speakers in the source data are disjoint from that of the train unit data. 

Unlike the baseline approach, the number of units in the proposed approach is not fixed before unit discovery. The acoustic units depend upon the number of clusters obtained by the initial clustering. For system 1, the number of units obtained are 170 and 112 for development and test datasets respectively. For system 2, the units are again clustered to 40 units. This number is tuned on development dataset by running for different values. Mel frequency cepstral coefficients extracted with a frame size of 25ms and frame shift of 10 ms are used as feature vectors. Different systems are built with static, delta cepstral features with mean subtraction, LDA with MLLT, FMLLR-based speaker adaptation. Kaldi toolkit \cite{povey2011kaldi} is used for feature extraction and training models. To train synthesis models, Merlin \cite{wu2016merlin}, an ANN based speech synthesis toolkit is used. As subjective evaluation is a time-consuming process, each team was allowed to submit a maximum of two systems. To shortlist the two systems, the multi-stimulus test with hidden reference and anchors (MUSHRA) \cite{itu2001bs} test was used. 

Previous challenges used minimal pair ABX discrimination \cite{schatz2013evaluating, schatz2014evaluating} score to evaluate the quality of the discovered units. As this Zerospeech 2019 is on speech synthesis, apart from ABX score, the primary subjective evaluation metrics include intelligibility in terms of character error rate (CER), speaker similarity, and overall quality of the synthesis in terms of mean opinion score (MOS). CER (0-1) is obtained by manually transcribing the synthesized audio and comparing the transcripts with the reference transcription. Speaker similarity (0-5) shows the similarity of the synthesized voice to that of the target speaker.

\begin{figure*}[t]
  \centering
  \includegraphics[width=1.0\linewidth]{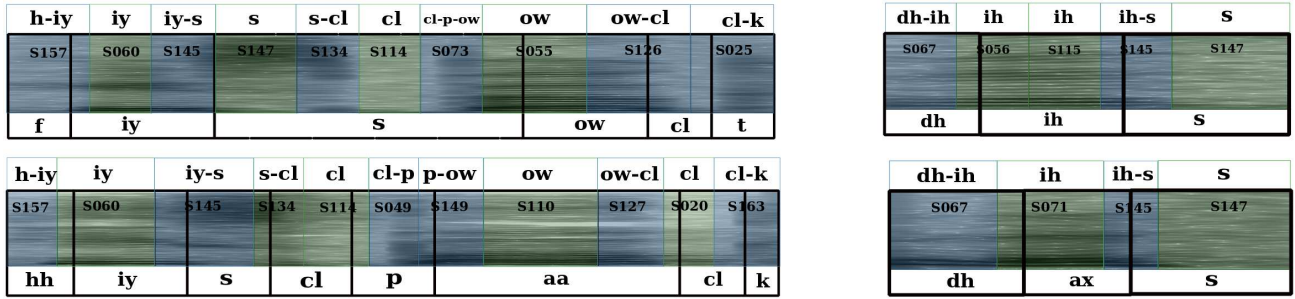}
  \caption{Segmentation of speech in terms of transient and steady-state acoustic units overlayed upon spectrograms. The utterances are \textit{"he spoke"} in the left side and \textit{"this"} in the right side}
  \label{fig:spec}
\end{figure*}

\section{Results and discussion}
\label{sec:results}

The baseline is a non-parametric sequential modelling technique proposed by Ondel et al. \cite{ondel2016variational}. It uses a phoneme-loop model, where each unit is modelled by an HMM. A Dirichlet process is assumed on the prior distribution of the units. The parameters of the HMMs means and covariance matrix are modelled using Normal-Gamma density, and the prior weights of the GMM and transition matrix of the HMM are modelled by Dirichlet distribution. Variational Bayes is used to infer the posterior distribution. A full fledged ASR is used as topline system. This system used transcription to perform supervised acoustic modelling. For speech synthesis, both baseline and topline systems uses Merlin toolkit.

Table~\ref{tab:result} shows the results on the baseline, topline and the two proposed approaches. The proposed approach is significantly better than the baseline system in terms of all the evaluation measures in both development and test languages. The number of acoustic units of the baseline system is 100 for both the datasets. The proposed system 1 had 170 and 112 units for development and evaluation datasets respectively, and the system 2 had 40 units for both the datasets. In-spite of using a small number of units, the system 2 had a bit-rate more than that of the baseline system. Since, each vowel is preceded and succeeded by rising and falling transients respectively, the proposed approach models each sonorant segment by three units, namely, a rising transient, a steady state, and a falling transient, giving rise to a higher number of symbols per unit of time.

The test language MOS score of the topline system is significantly higher than that of the baseline and the proposed systems. This is in contrast to the scores of development language. This difference can be attributed to the fact that the grapheme-phoneme correspondence in the Indonesian language, being the Malay language, is almost perfect \cite{lee2011acquisition}, as opposed to that of English. Because of  allophonic variations in English, the MOS scores on the proposed approach is better than that of the supervised acoustic modelling and synthesis modelling. This inconsistency due to allophonic variations is also reflected in the objective evaluation measure ABX. The ABX measures of the proposed approach and the top-line are close in English, whereas the measure is much better for top-line in the case of Indonesian. 

AUD problem can also be compared to that of children's language acquisition \cite{lindblom2000developmental}. During language acquisition, the syllable acts as the basic units that a child learns by continuous exposure. After the basic syllabic units are learnt, a child starts learning  complex patterns that make up continuous speech. The proposed approach uses a similar technique for AUD, wherein first the training is confined to syllable-like units. Then subsequently, the models are used to bootstrap continuous speech training.

The final acoustic units consist of steady state and transient units. The steady-state regions include steady-state sonorous sounds and fricatives, and the consonant transition from and to steady-state regions are modelled as transient regions. Figure~\ref{fig:spec} shows the segmentation results of system1 overlayed on spectrograms on two different utterances uttered by two different speakers. The top two pictures correspond to the target speaker and the bottom pictures correspond to two different train speakers. In each picture, three labels panes are shown. As the phoneme alignment is not known, an alignment is obtained from the models trained on TIMIT corpus. The bottom pane shows this segmentation. The middle pane shows the cluster ID. The top pane shows the phonetic mapping of the acoustic units to phonemes. In the figure, the blue regions correspond to transition and green regions correspond to steady-state. The figure shows that the AU transciption is better than phoneme transciption obtained from  supervised training using TIMIT dataset. Even though the transients across different speakers have identical acoustic units across speakers, the symbols of the vowel-steadty state regions are not identical. This can mean that, in the absence of supervised training, the same vowel belonging to different speakers with high variability in fundamental frequency may not be modelled as a single unit. Even the supervised system transcribed the vowels in an utterance uttered by two different speakers as different. It is a well known fact that the recognition accuracy of the consonants is very low for any phoneme recognizer. As the proposed approach models the consonants with context exclusively, the consonantal transients (CV and VC) are recognized with high recognition accuracy, as observed in Figure~\ref{fig:spec}. 

Modelling of speech as transient and steady-state acoustic units is in accordance with other linguistic studies \cite{massaro1975_77} that show that the steady-state vowels and CV, VC transients are basic units of speech perception. Massaro, in his studies, has classified the CV transients as stop transient, nasal transient, fricative transient. The clustered units in system 2 indeed segregates similar kind of transients into one cluster. Even though for the synthesis task, the results from the subjective evaluation is better for system 1, it is observed that for other task such as spoken term detection, the system 2 outperforms system 1.

\section{Conclusions}
\label{sec:conclusion}
In the absence of transcribed data to train acoustic models for speech recognition and speech synthesis, the syllabic structure present in speech is valuable information. This information can not only be used as initial segments but can also be used to discover units which are perceptual. The results show that such units, when modelled using simple GMM-HMM framework, can achieve good synthesis quality and speaker similarity with reduced bit-rate.

\bibliographystyle{IEEEtran}

\bibliography{refs}

\end{document}